# A multi-layer refined network model for the identification of essential proteins

Haoyue Wang, Li Pan*, Bo Yang, Junqiang Jiang and Wenbin Li*

**Abstract**—The identification of essential proteins in protein-protein interaction networks (PINs) can help to discover drug targets and prevent disease. In order to improve the accuracy of the identification of essential proteins, researchers attempted to obtain a refined PIN by combining multiple biological information to filter out some unreliable interactions in the PIN. Unfortunately, such approaches drastically reduce the number of nodes in the PIN after multiple refinements and result in a sparser PIN. It makes a considerable portion of essential proteins unidentifiable. In this paper, we propose a multi-layer refined network (MR-PIN) that addresses this problem. Firstly, four refined networks are constructed by respectively integrating different biological information into the static PIN to form a multi-layer heterogeneous network. Then scores of proteins in each network layer are calculated by the existing node ranking method, and the importance score of a protein in the MR-PIN is evaluated in terms of the geometric mean of its scores in all layers. Finally, all nodes are sorted by their importance scores to determine their essentiality. To evaluate the effectiveness of the multi-layer refined network model, we apply 16 node ranking methods on the MR-PIN, and compare the results with those on the SPIN, DPIN and RDPIN. Then the predictive performances of these ranking methods are validated in terms of the identification number of essential protein at top100 - top600, sensitivity, specificity, positive predictive value, negative predictive value, F-measure, accuracy, Jackknife, ROCAUC and PRAUC. The experimental results show that the MR-PIN is superior to the existing refined PINs in the identification accuracy of essential proteins.

## 1 Introduction

PROTEINS are the main undertaker of biological life activities and are the most important components of living organisms with very important biological functions. The essential proteins are closely related to the life activities of biological cells, and their absence will lead to the inability of organisms to survive [1]. The identification of essential proteins is conducive to understand the minimum requirements for cell survival and development, which helps to find drug targets and prevent diseases [2], [3]. Proteins participate in various aspects of life activities such as signal transmission, gene expression regulation through interaction with each other, thus forming a protein-protein interaction network (PIN).

So far, with the rapid development of high-throughput technology, a large amount of PIN data has been accumulated. In order to identify more essential proteins from PIN, some researchers have proposed node importance ranking methods, that is, using different biological information to calculate the importance score of proteins [4]-[32], while other researchers have used prediction methods based on machine learning or deep learning, that is, first, an effective classification model is trained by using the biological information and labels of some proteins, and then the essentiality of the remaining proteins are judged by this model [33]-[35].

In order to better understand the similarities and differences among the node importance ranking methods, we further classify these methods into the following two categories. The first type is the network-based method, which calculates the importance scores of proteins in a specific PIN (static PIN, weighted PIN, refined PIN) to identify essential proteins. In general, most network-based centrality methods [6]-[16] have been proposed based on static PIN (static PIN is a network composed of raw dataset of protein-protein interactions), such as, degree centrality (DC) [6], local average connectivity centrality (LAC) [7], closeness centrality (CC) [13], betweenness centrality (BC) [14], eigenvector centrality (EC) [16], etc. Unfortunately, the PIN obtained from high-throughput biological experiments have been found to contain a lot of noise [17]. Limited by the poor quality of the underlying PIN data, these methods have shown high false-positive and false-negative results in measuring the importance of proteins. In order to overcome this problem, in recent years, some researchers have weighted the interaction of protein pairs in PIN by other biological information of proteins to identify more essential proteins, namely, the methods based on weighted PIN [18]-[20]. For example, Lei et al. [19] used RNA sequences and GO annotations to judge the co-expression degree and functional similarity of protein pairs to build a more reliable weighted network and proposed the RSG method by combining the weighted edge clustering coefficient and subcellular localization information. Li et al. [20] constructed a weighted PIN by integrating the neighborhood density, gene expression profile and subcellular location information, and proposed the TGSO method by introducing protein orthologous score and iterative method. In addition, some researchers are committed to removing the unreliable interactions in the original PIN by combining other biological information with PIN, so as to obtain

- *H. Y. Wang, L. Pan, B. Yang, J. Q. Jiang and W. B. Li are with Hunan Institute of Science and Technology, Yueyang, Hunan, China. E-mail: { lipan@hnist.edu.cn, wenbin_lii@163.com}.*
- **Corresponding author*

a finer PIN and improve the performance of the node importance ranking methods, namely, the methods based on refined PIN [21]-[24]. For example, Xiao et al. [21] combined PIN with gene expression profile, removed some unreliable interactions by determining whether the protein pairs were expressed at the same time, and proposed a dynamic PIN (DPIN). Subsequently, Li et al. [22], on their basis, considered the spatial properties of proteins and further removed the unreliable interactions by determining whether the protein pairs were located in the same subcellular compartment, thus proposed the refined dynamic PIN (RDPIN). The second type is the data fusion method based on multiple biological information (such as topological information of proteins in PIN, gene expression profile, subcellular localization information, complex, orthologous information, etc.) that uses multiple biological information of proteins to calculate their importance scores [25]-[31]. For example, Li [25] et al. and Tang [26] et al., proposed PeC and WDC methods by integrating the degree of co-expression between protein pairs in gene expression profiles and the edge clustering coefficients of their interactions. Li et al. [27] pointed out that the protein occurred multiple times in complex is more likely to be essential and proposed the UC method which coupled the frequency of protein occurrences in complexes and the edge clustering coefficient. Qin et al. [28] proposed the LBCC method by combining the local density of proteins, betweenness centrality and in-degree centrality of complex. Zhong et al. [31] used a dynamic threshold method to binarize gene expression values and proposed the JDC method in order to combine the co-expression states and edge clustering coefficients of protein pairs at multiple times.

Studies have proved that biological information can filter some unreliable interactions between proteins in PIN, so as to obtain refined PIN [25], [26], which can effectively improve the identification accuracy of essential proteins. However, when multiple biological information is progressively used to filter the interactions in a PIN, the interactions in the PIN are greatly reduced after repetitive refinements. For the DIP dataset, in the DPIN, there are about 9% of essential proteins that lose their interactions by one refinement. But in the RDPIN, over 27% of essential proteins lose their interactions due to two refinements. If the PIN is thirdly refined by protein complex informati on, then over 50% of essential proteins will lose their inter actions. Hence, multiple refinements will result in a very sparse PIN, and thus affect the identification accuracy of essential proteins.

To address this problem, we propose a multi-layer refined network model (MR-PIN) for the identification of essential proteins. First, four refined networks are constructed by respectively integrating gene expression profile, subcellular localization information, protein complex or orthologous information into static PIN (SPIN). Then, the proteins are scored in each network layer by using node importance ranking methods, and the importance score of each node is calculated in terms of the geometric mean of its scores in all layers. Finally, we sort all nodes by their importance scores to determine the essentiality of proteins. Different from the existing PINs constructed by multiple refinements, we construct a multi-layer network that consists of four once-refinement heterogeneous networks. The two advantages of this treatment are that (1) the interactions of essential proteins are saved as much as possible in each network layer and (2) different biological features are kept in their own network layers instead of being mixed in a single aggregation network (see Fig.1 and Fig. 2). All of these are beneficial to the identification of essential proteins.

To evaluate the effectiveness of the multi-layer refined network model, we apply 16 node importance ranking methods (LAC, NC, DMNC, LID, DC, EC, PR, BC, CC, TP, CLC, PeC, WDC, LBCC, UC, JDC) on the MR-PIN, and compare the results with those on SPIN, DPIN and RDPIN. Then the predictive performances of these methods are validated in terms of the identification number of essential protein at top100 - top600, sensitivity, specificity, positive predictive value, negative predictive value, F-measure, accuracy, jackknifing method, ROCAUC and PRAUC. The experimental results show that the MR-PIN is superior to the existing PINs in the identification accuracy of essential proteins.

## 2 Methods

### 2.1 Construction of the MR-PIN

First, we denote a PIN as an undirected graph, and refer to an original PIN that does not incorporate any biological information as a static protein-protein interaction network (SPIN), denoted by $G = (V, E)$, where $V$ is the set of all proteins and E is the set of all protein interactions. Then, based on gene expression profiles, subcellular localization information, complexes and orthologous information, the following four once-refinement PINs containing different biological information were considered for construction: the 1st network layer ($G_T$), the 2nd network layer ($G_S$), the 3rd network layer ($G_M$), and the 4th network layer ($G_C$). The overall idea of the multi-layer refined network model (MR-PIN) is shown in Figure 2. The four layers are the induced subgraphs of the static PPI network from the different biological perspective, and therefore do not increase any other interactions relative to the static network.

(1) Construction of the 1st network layer.

Since the interactions of proteins in cells are not static but vary with time, constructing a time-dependent PIN can better reflect the temporal properties of PIN. Protein is translated from its corresponding mRNA, the dynamic characteristic of PIN can be indirectly reflected from gene expression level data, and is mainly reflected in that gene is expressed at some moments and not at others. Based on the gene expression level data, first, we calculated the activity of each protein, and the activity threshold $\tau_i$ for protein $v_i$ was calculated using the following equation [5], [36].

$$\tau_i = \mu_i + p\sigma_i \tag{1}$$

Where $\mu_i$ denotes the mean of the 36 time-point gene ex-

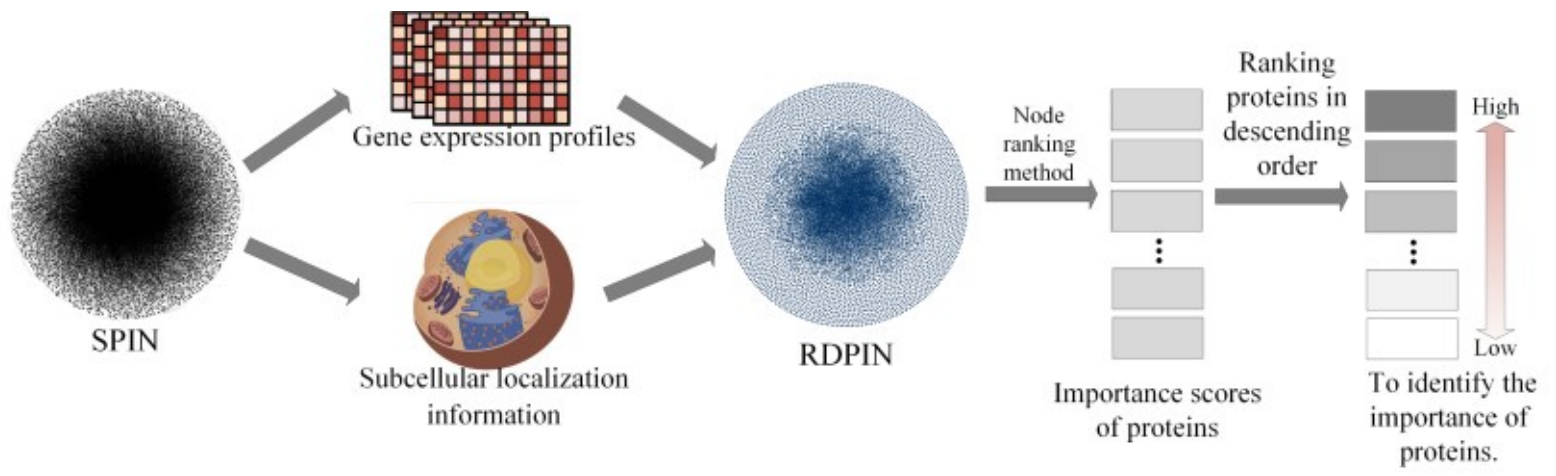

Figure 1. Construction of RDPIN and identification of essential proteins.

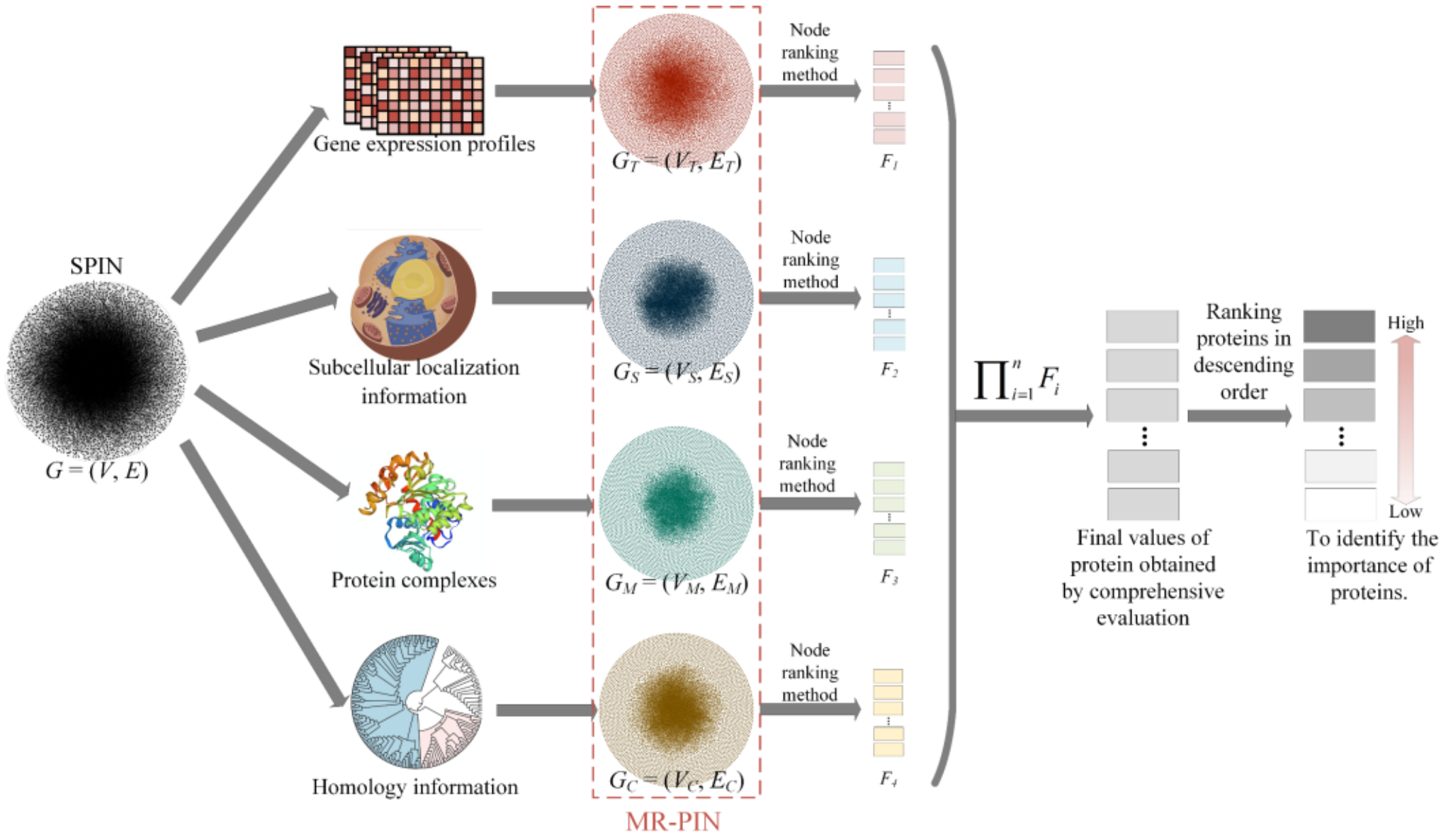

Figure 2. The overall idea of the multi-layer refined network model (MR-PIN). Where $G_T$, $G_S$, $G_M$ and $G_C$ are the induced subgraphs of the static network (SPIN), which constitute the multi-layer heterogeneous network.

pression level values for $v_i$, $\sigma_i$ is the standard deviation of the gene expression level values for $v_i$, and $p$ is an adjustable threshold, generally $p \in (0,3]$. Let $e_{ik}$ represents the gene expression value of $v_i$ at time-point $t_k$. If $e_{ik}$ is greater than $\tau_i$, $v_i$ will be expressed at time-point $t_k$, and let $e_{ik} = 1$. Secondly, the sequence $G_1, G_2, ..., G_{36}$ of subnets for time-point $t_1, t_2, ..., t_{36}$ is derived from $G$. The idea is, for the subnet $G_k = (V_k, E_k)$ at time-point $t_k$, there are $V_k = V$, for $(v_i, v_j) \in E$, if $v_i(e_{ik}) = v_j(e_{jk}) = 1$ at time-point $t_k$, $(v_i, v_j) \in E_k$. Thirdly, the 36 time-point subnets are aggregated into a single-layer network $G_T = (V_T, E_T)$, where

(i) $V_T = V$, and

(ii) $E_T = \{(v_i, v_j) \mid (v_i, v_j) \in E \cap \exists t_k, (v_i, v_j) \in E_k\}$.

If two proteins interacting in the static network are co-expressed at some time-point, their interaction is more likely to be real from the biological perspective of the gene expression level data and to be retained in the 1st layer. On the contrary, their interaction is abandoned.

(2) Construction of the 2nd network layer.

The spatial characteristic of protein is that they have different subcellular compartments and play different roles and importance in cellular activities. Among them, the nucleus is crucial in cell growth, differentiation and metabolism, and cells without a nucleus cannot survive for long. Furthermore, it has been shown that proteins, even essential proteins have the most widespread distribution in the nucleus [30], so we consider constructing a space-related network. If protein $vi$ occurs in the nucleus, we denote $Nu(v_i) = 1$, otherwise $Nu(v_i) = 0$. The spatial subnetwork (the 2nd layer) $G_s = (V_s, E_s)$ can be defined as:

(i) $V_s = V$, and

(ii) $E_s = \{(v_i, v_j) \mid (v_i, v_j) \in E$ and $Nu(v_i) = Nu(v_j) = 1\}$.

If the two proteins $v_i$ and $v_j$ that interact in the static network are co-localized in the nucleus, their interaction is more likely to be real from the biological perspective of the subcellular compartment and to be retained in the 2nd layer.

(3) Construction of the 3rd network layer.

It has been shown that PINs have modular characteristics, and that the necessity of a protein depends not only on the protein itself but also on the functional module in which the protein is located, i.e., the protein complex [5], [37], [38]. Furthermore, proteins in complexes are more likely to be essential than proteins do not present in the complex [27]. Thus, protein complexes contribute to the identification of essential proteins. In this paper, we combined protein complexes and PIN to obtain PIN containing functional modularity information. For protein $v_i$, we denote by $C(v_i)$ the set of complexes in which protein $v_i$ appears. Let $G_M = (V_M, E_M)$ be the modular subnetwork (the 3rd layer), where

(i) $V_M = V$,

(ii) $E_M = \{(v_i, v_j) \mid (v_i, v_j) \in E$ and $C(v_i) \cap C(vj) \neq \varnothing\}$.

That is, if two proteins interacting in the static network are present in the at least a same complex, then their interactions are more likely to be real from the biological perspective of protein complexes and to be retained in the 3rd layer.

(4) Construction of the 4th network layer.

If two or more protein sequences share the same ancestor, they are said to be homologous, and the homologous part of the sequence is said to be conserved. It has been pointed out that essential proteins are conserved [18], so they evolve at a much slower rate than non-essential proteins, in other words, essential proteins are more stable. In this paper, we consider combining orthologous information of proteins with PIN to filter out some unreliable interactions and obtain a PIN containing protein conservative characteristics. For protein $v_i$, let $O(v_i)$ represent the set of reference organisms in which at least an orthologous protein pair including $v_i$ occurs. Let $G_C = (V_C, E_C)$ be the conservative subnetwork (the 4th layer), where

(i) $V_C = V$,

(ii) $E_C = \{(v_i, v_j) \mid (v_i, v_j) \in E$ and $|O(v_i)| \geq q$ and $|O(v_j)| \geq q$ and $O(v_i) \cap O(v_j) \neq \varnothing\}$.

If two proteins interacting in the static network are present in at least one of the same reference genomes, and their orthologous scores $|O(v_i)|$ and $|O(v_j)|$ are greater than the threshold $q$, then their interactions are more likely to be real from the biological perspective of orthologous relationships and to be retained in the 4rd layer.

### 2.2 Node importance ranking for the MR-PIN

In this paper, we used node importance ranking methods to measure the importance of proteins in the MR-PIN, the steps are as follows: (i) Firstly, the node importance ranking methods used in this paper take a single-layer network (typically represented by an adjacency matrix) as input, and output importance scores of all protein in the network. For the 4-layer network ($G_T$, $G_S$, $G_M$, $G_C$), we take each layer as input respectively, and calculate four importance scores of each protein. But it's worth noting that in different refined subnetworks, the dimensions of the importance score of the same protein are different. Furthermore, sometimes there are more zero values when using node importance ranking methods to score proteins in refined networks, which will affect the subsequent comprehensive evaluation of proteins. Therefore, we need to normalize the scores of proteins in each refined network and add a small deviation of 0.001 to the normalization so that the centrality values in each network are scaled to (0,1]. The formula for scaling the centrality values is shown below.

$$F_i = \frac{f_i - \min(f_i) + 0.001}{\max(f_i) - \min(f_i) + 0.001} \quad (2)$$

Where $f_i$ is the importance scores for all proteins in the $i$th refined network, $\min(f_i)$ is the minimum value in $f_i$ and $\max(f_i)$ is the maximum value in $f_i$. (ii) Secondly, after obtaining the four importance scores of the same dimension of each protein, we calculated the comprehensive scores for each protein as its final score in the MR-PIN, and the higher final score of the protein, the more important the protein is. The formula is as follows.

$$protein_scores = \prod_{i=1}^{n} F_i \quad (3)$$

Where, $n$ is the number of refinement networks, in this paper, $n$=4, $F_i$ is the importance scores of proteins after normalization in the $i$th refined network. (iii) Finally, we sort comprehensive scores of all proteins in descending order, then take the top K ranked proteins as essential proteins (K=1130 for DIP and K=1199 for BioGRID) and calculate the prediction accuracy.

## 3 Experiment and Discussion

### 3.1 Datasets

**Protein-protein interaction datasets:** In this study, we used the Saccharomyces cerevisiae datasets that are now widely used to test various approaches for the discovery of essential proteins. The two protein interaction networks in the paper were downloaded from the datasets DIP [39] and BioGRID [40], which contain 4,746 and 5,616 proteins respectively.

**Essential proteins:** The essential proteins are collected from the following datasets [41], [42], [43]: DEG, MIPS, SGD, SGDP. According to refs [11], [23] and [29], a protein is viewed as essential if it is labeled as essential protein in at least one database. From the above four databases, 1,130 essential proteins are collected for the PPI network constructed by the DIP dataset and 1,199 essential ones for that by the BioGRID dataset. In addition, a small number of unknown proteins were classified as non-essential proteins.

**Gene expression profiles:** The yeast gene expression data is obtained from the NCBI GEO database. GSE3431 gene expression profile is used in this paper [44], which contains 6,777 proteins, covering three consecutives metabolic cycles, and each cycle containing 12 time point gene expression values. In addition, the gene expression level data is normally distributed and has been normalized. In DIP and BioGRID, 98.6% and 94.3% of the proteins are in the gene expression profile, respectively.

**Subcellular localization information:** The subcellular localization information of proteins is downloaded from the COMPARTMENTS database [45], which contains 11 subcellular locations: cytoskeleton, golgiapparatus, cytosol, endosome, mitochondrion, plasma membrane, nucleus, extracellular space, vacuole, endoplasmmic, reticulum, peroxisome. Among them, the nucleus contains the most essential proteins, with 63.9% of the essential proteins present in the nucleus in both DIP and BioGRID.

**Protein complex:** The protein complexes data is from CM270, CM425, CYC408 and CYC428 database [46], [47], [48], which can be obtained from [49], including 745 protein complexes, covering a total of 2,167 proteins.

**Orthologous information:** Information on orthologous proteins is taken from Version 7 of the InParanoid database (an orthologous database), which contains a collection of paired comparisons between 100 whole genomes (99 eukaryotes and 1 prokaryote) [50], [51].

### 3.2 Experimental analysis on DIP dataset

In order to comprehensively compare the performance of MR-PIN, we used 16 Node importance ranking methods to study their improvement rates in MR-PIN. Among them, there are 11 network-based centrality methods, which use only one kind of biological information, namely protein-protein interaction network: DC [6], LAC [7],

NC [8], DMNC [9], TP [10], LID [11], CLC [12], CC [13], BC [14], PR [15], EC [16], and 5 approaches using more than two kinds of biological information: PeC [18], WDC [19], UC [20], LBCC [21], JDC [24].

### 3.2.1 Analysis of the identification number of essential proteins

To demonstrate the superiority of MR-PIN over existing network refinement methods, we tested 16 different node importance ranking methods and compared MR-PIN with SPIN, DPIN, and RDPIN for the identification number of essential proteins at top100, top200, top300, top400, top500, and top600, as shown in Figure 3 (the results of MR-PIN in the figure are the optimal results of each method). It can be seen that the MR-PIN model has the most obvious improvement effect on these methods. Compared with RDPIN, the best existing refinement network, 11 network-based centrality methods can generally improve the identification accuracy of essential proteins by at least 10.51% at top600, among which, some methods have very remarkable improvement: for example, compared with RDPIN, CLC method and PR method can improve the identification accuracy of essential proteins by 31.61% and 25.38%, respectively. At the same time, 5 node importance ranking methods containing more than two biological information can also improve the identification accuracy of essential proteins by 8.94% to 13.93%. It is worth noting that on the MR-PIN model, the identification accuracies of essential proteins of 12 methods are more than 66.67% at top600, i.e., LAC, NC, DMNC, LID, DC, EC, PR, CC, TP, CLC, WDC and UC. Among them, the DC method and the TP method have the highest identification numbers of essential proteins at top600, which is about 69.67%. It is worth noting that the identification number of essential proteins of JDC method achieved 99 at top100, which is extremely a high identification accuracy. Therefore, the MR-PIN model can improve the performance of the node importance ranking methods.

### 3.2.2 Validated by using the jackknifing method

The jackknifing method has been used in [18], [22], [29], [52] to evaluate the effectiveness of essential protein discovery methods. In this paper, the jackknifing method is used to compare the performance of node ranking methods on the proposed multi-layer network with those on other refined networks. In jackknifing method, proteins are ranked by each essential protein discovery method and the cumulative count of essential proteins is plotted. The x-axis represents the number of ranked proteins in each method and the Y-axis is the cumulative count of true essential proteins with respect to the ranked proteins moving left to right. As shown in Figure 4, it can be seen that the overall performance of all 16 methods on the MR-PIN model is optimal, compared with that on SPIN, DPIN and RDPIN. Among them, the improvements of network-based centrality methods are more obvious. All these prove that the MR-PIN model is superior to the existing network refinement methods, and each node importance ranking method can identify more essential proteins in this model.

### 3.2.3 Validate by ROCAUC and PRAUC

To further evaluate the significance of the MR-PIN, first, we calculated the areas under the receiver operating characteristic curves (ROCAUC) of 16 node importance ranking methods on the SPIN, DPIN, RDPIN and MR-PIN. Secondly, because the number of essential and non-essential proteins is unbalanced, the non-essential proteins are about three times as many as the essential proteins, so the identification of essential proteins is a matter of sample imbalance. In fact, we are more concerned with how many essential proteins can be identified. Therefore, we also calculated the area under the precision-recall curve (PRAUC) of each node importance ranking method. As shown in Table 1, it can be seen that on MR-PIN, the values of ROCAUC and PRAUC of different methods have been improved to varying degrees, and some of them have prominent improvement rates. For example, compared with SPIN, CC method can improve the values of ROCAUC and PRAUC by 21.94% and 65% on MR-PIN, respectively. Even compared with the existing optimal refinement network RDPIN, PR method can also improve the values of ROCAUC and PRAUC by 11.22% and 29.21% on MR-PIN, respectively. All of this prove the validity of MR-PIN.

Table 1. Comparison of ROCAUC and PRAUC of 16 node importance ranking methods on SPIN, DPIN, RDPIN, and MR-PIN on DIP dataset (ROCAUC/PRAUC).

| Methods | SPIN | DPIN | RDPIN | MR-PIN |
|---|---|---|---|---|
| LAC | 0.690/0.484 | 0.682/0.503 | 0.673/0.517 | **0.714/0.568** |
| NC | 0.689/0.478 | 0.681/0.496 | 0.672/0.506 | **0.713/0.554** |
| DMNC | 0.676/0.421 | 0.674/0.451 | 0.667/0.465 | **0.713/0.557** |
| LID | 0.691/0.488 | 0.683/0.513 | 0.675/0.528 | **0.712/0.557** |
| DC | 0.687/0.411 | 0.692/0.435 | 0.718/0.497 | **0.785/0.592** |
| EC | 0.659/0.413 | 0.673/0.450 | 0.711/0.493 | **0.757/0.574** |
| PR | 0.672/0.388 | 0.677/0.400 | 0.704/0.445 | **0.783/0.575** |
| BC | 0.638/0.350 | 0.644/0.363 | 0.674/0.433 | **0.723/0.534** |
| CC | 0.638/0.360 | 0.645/0.372 | 0.708/0.456 | **0.778/0.594** |
| TP | 0.662/0.387 | 0.676/0.419 | 0.717/0.497 | **0.785/0.594** |
| CLC | 0.665/0.376 | 0.668/0.418 | 0.662/0.439 | **0.709/0.530** |
| PeC | 0.638/0.431 | 0.660/0.435 | 0.690/0.470 | **0.689/0.518** |
| WDC | 0.688/0.486 | 0.695/0.483 | 0.700/0.516 | **0.718/0.554** |
| LBCC | 0.687/0.468 | 0.671/0.469 | 0.657/0.477 | **0.695/0.545** |
| UC | 0.692/0.483 | 0.683/0.502 | 0.673/0.509 | **0.714/0.570** |
| JDC | 0.697/0.514 | 0.684/0.522 | 0.674/0.529 | **0.707/0.575** |

### 3.2.4 Validate by six evaluation indicators

To further evaluate the overall performance of MR-PIN and the accuracy of essential protein identification, we used the following six evaluation metrics: sensitivity (SN), specificity (SP), positive predictive value (PPV), negative predictive value (NPV), F-measure (FM), and accuracy (ACC). The top 1130 proteins (1130 is the number of essential proteins for DIP) after the descending order of each centrality metric value were assumed to be essential proteins, and the calculation formulas are as follows, where TP is the correctly predicted essential protein, FP stands for the incorrectly predicted essential protein, TN refers to the correctly predicted non-essential protein, and FN represents the incorrectly predicted non-essential protein.

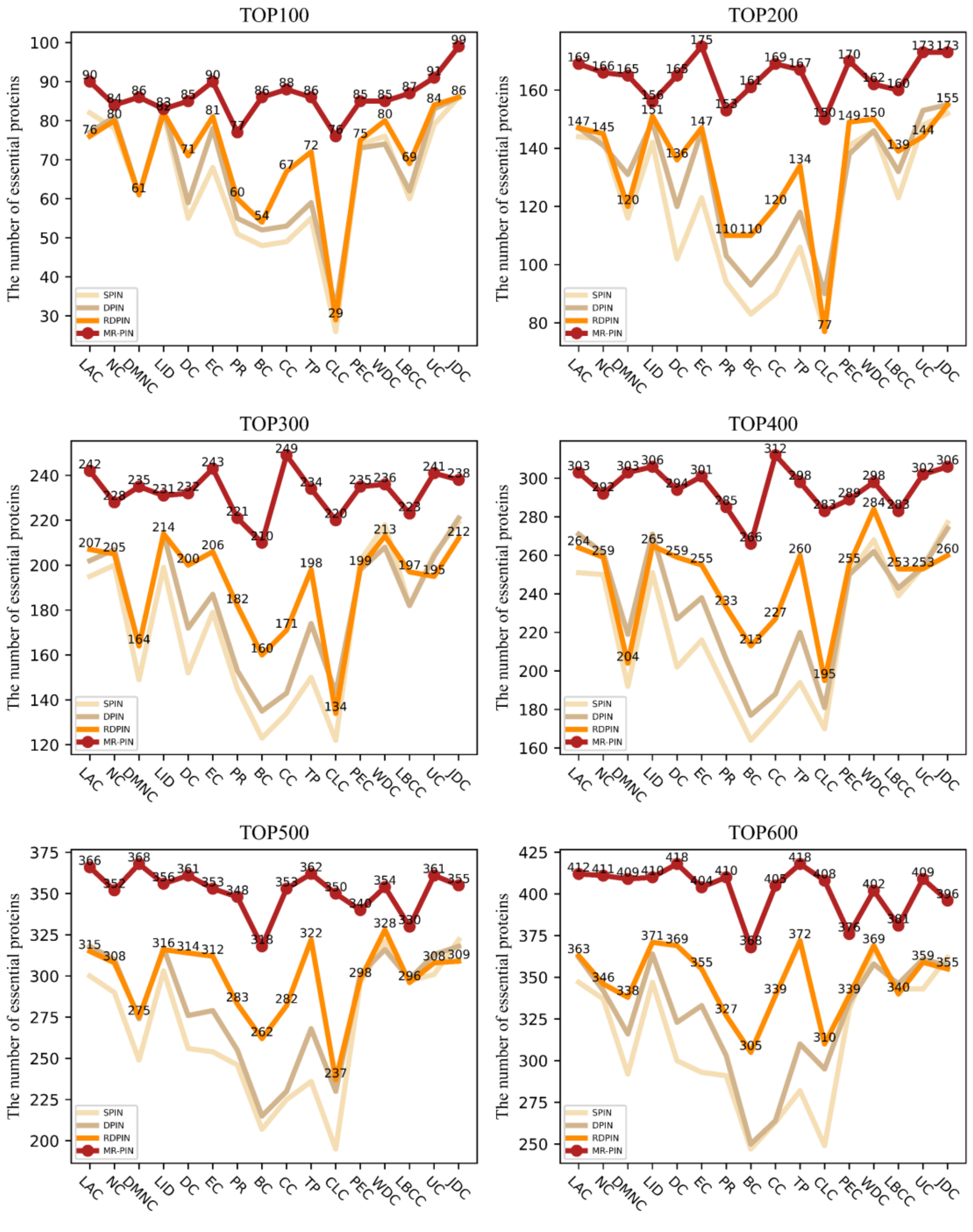

Figure 3. Comparison of the number of essential proteins identified by 16 node importance ranking methods on SPIN, DPIN, RDPIN, and MR-PIN on DIP dataset.

$$SN = TP / (TP + FN) \quad (4)$$

$$SP = TN / (FP + TN) \quad (5)$$

$$PPV = TP / (TP + FP) \quad (6)$$

$$NPV = TN / (TN + FN) \quad (7)$$

$$FM = 2 \cdot SN \cdot PPV / (SN + PPV) \quad (8)$$

$$ACC = (TP + TN) / (TP + TN + FP + FN) \quad (9)$$

Table 2 shows the comparison of six evaluation indicators of 16 node importance ranking methods on MR-PIN and three existing networks (SPIN, DPIN, RDPIN). It can be seen that the six evaluation indexes of these methods and the identification number of essential proteins at top1130 are all the best on the MR-PIN, which undoubtedly proves that the MR-PIN can improve the identification accuracy of essential proteins of the node importance ranking methods.

### *3.2.5 Analysis of feasibility and ablation experiments of MR-PIN*

In order to prove the feasibility of the method proposed in this paper, as shown in Table 3, first, we listed the top600, top1130, PRAUC and ROCAUC values of the 10 centrality methods (because network-based centrality methods are more sensitive to the topological structure of PIN) in once-refinement networks (lines 2-5 in Table 3). It can be seen that it is feasible to use gene expression profiles, subcellular localization information, protein complex and orthologous information to once refine the network respectively to improve the identification accuracy of essential proteins. Compared with the SPIN, their performance has been improved to different degrees. However, among the four once-refinement networks, the $G_T$ refined with gene expression profile had the least improvement effect, while the $G_C$ refined with orthologous information had the greatest improvement effect. Secondly, we further used two different once-refinement networks to form two-layer heterogeneous networks and used the method in this paper to comprehensively score proteins. According to the experimental results of 10 centrality methods (lines 6-11 in Table 3), it can be seen that the two-layer heterogeneous network has better performance than the single-layer refined network. Therefore, it is feasible to use multi-layer network method to identify more essential proteins. As can be seen from Table 3, the overall performance of the MR-PIN model is optimal. Furthermore, in order to validate the importance of each of the four once-refinement networks on the MR-PIN, we conducted an ablation experiment on the four once-refinement networks on the MR-PIN (lines 12-15 in Table 3). In the model without $G_C$, the average difference of ROCAUC values of the 10 centrality methods relative to MR-PIN was 2%, which was the largest among all the models in the ablation experiment, suggesting that $G_C$ played the largest role on the MR-PIN overall. On the contrary, in the model without $G_T$, the average difference of ROCAUC values of the 10 centrality methods relative to MR-PIN was 0.38%, which was the smallest among all the models in the ablation experiment, indicating that the role of $G_T$ on the MR-PIN was the smallest.

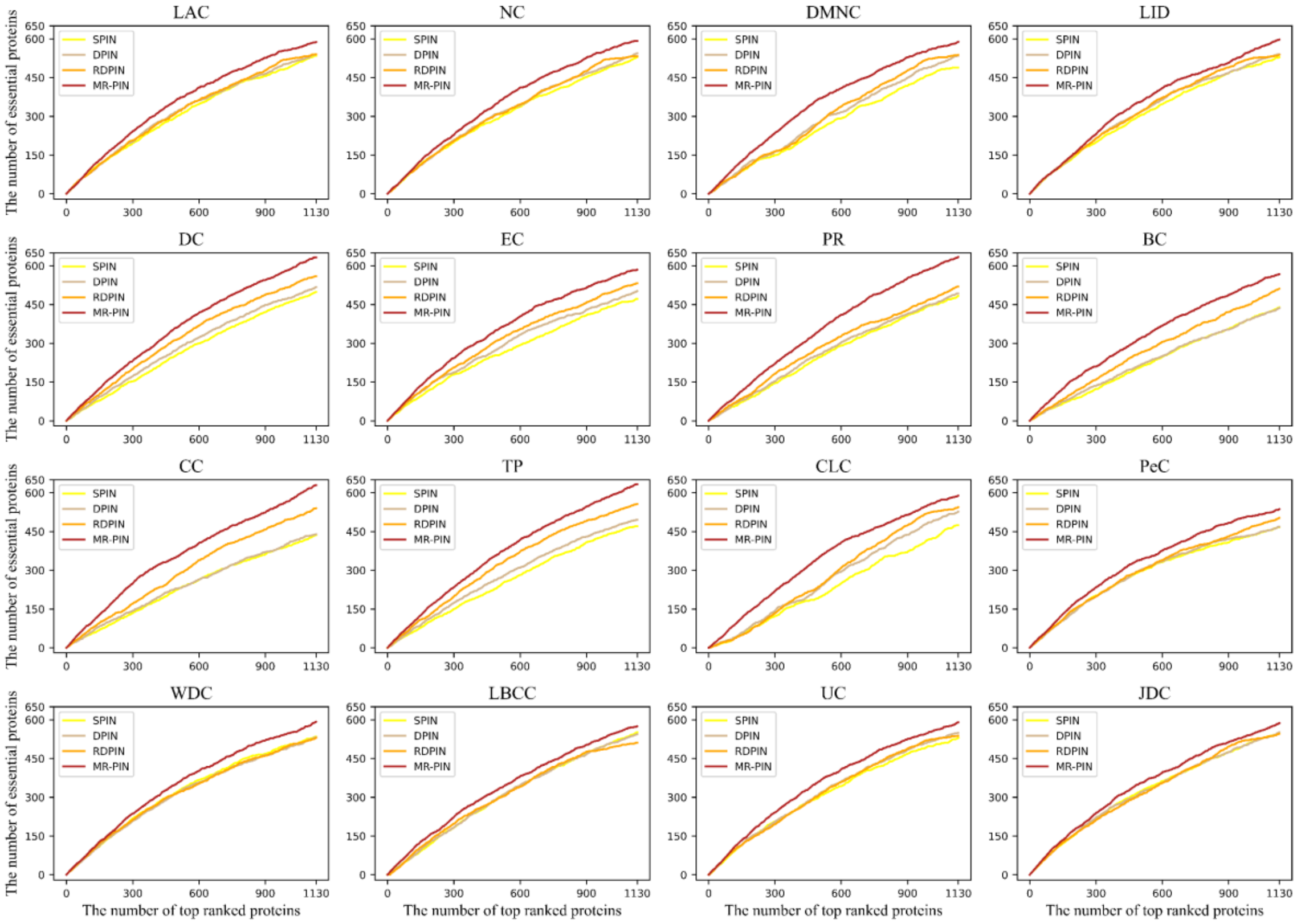

Figure 4. 16 node importance ranking methods are validated by the jackknifing method on DIP dataset.

Table 2. Comparison of six evaluation indicators for 16 node importance ranking methods on DIP dataset.

| Centrality | PIN | SN | SP | PPV | NPV | FM | ACC | TOP1130 |
|---|---|---|---|---|---|---|---|---|
| LAC | SPIN | 0.4735 | 0.8355 | 0.4735 | 0.8355 | 0.4735 | 0.7493 | 535 |
| | DPIN | 0.4779 | 0.8368 | 0.4779 | 0.8368 | 0.4779 | 0.7514 | 540 |
| | RDPIN | 0.4779 | 0.8368 | 0.4779 | 0.8368 | 0.4779 | 0.7514 | 540 |
| | MR-PIN | 0.5195 | 0.8498 | 0.5195 | 0.8498 | 0.5195 | 0.7712 | 587 |
| NC | SPIN | 0.4681 | 0.8338 | 0.4681 | 0.8338 | 0.4681 | 0.7467 | 529 |
| | DPIN | 0.4805 | 0.8377 | 0.4805 | 0.8377 | 0.4805 | 0.7526 | 543 |
| | RDPIN | 0.4717 | 0.8349 | 0.4717 | 0.8349 | 0.4717 | 0.7484 | 533 |
| | MR-PIN | 0.5239 | 0.8512 | 0.5239 | 0.8512 | 0.5239 | 0.7733 | 592 |
| DMNC | SPIN | 0.4327 | 0.8227 | 0.4327 | 0.8227 | 0.4327 | 0.7299 | 489 |
| | DPIN | 0.4735 | 0.8355 | 0.4735 | 0.8355 | 0.4735 | 0.7493 | 535 |
| | RDPIN | 0.4761 | 0.8363 | 0.4761 | 0.8363 | 0.4761 | 0.7505 | 538 |
| | MR-PIN | 0.5204 | 0.8501 | 0.5204 | 0.8501 | 0.5204 | 0.7716 | 588 |
| LID | SPIN | 0.4673 | 0.8335 | 0.4673 | 0.8335 | 0.4673 | 0.7463 | 528 |
| | DPIN | 0.4779 | 0.8368 | 0.4779 | 0.8368 | 0.4779 | 0.7514 | 540 |
| | RDPIN | 0.4761 | 0.8363 | 0.4761 | 0.8363 | 0.4761 | 0.7505 | 538 |
| | MR-PIN | 0.5283 | 0.8526 | 0.5283 | 0.8526 | 0.5283 | 0.7754 | 597 |
| DC | SPIN | 0.4416 | 0.8255 | 0.4416 | 0.8255 | 0.4416 | 0.7341 | 499 |
| | DPIN | 0.4584 | 0.8308 | 0.4584 | 0.8308 | 0.4584 | 0.7421 | 518 |
| | RDPIN | 0.4947 | 0.8421 | 0.4947 | 0.8421 | 0.4947 | 0.7594 | 559 |
| | MR-PIN | 0.5602 | 0.8626 | 0.5602 | 0.8626 | 0.5602 | 0.7906 | 633 |
| EC | SPIN | 0.4177 | 0.8180 | 0.4177 | 0.8180 | 0.4177 | 0.7227 | 472 |
| | DPIN | 0.4442 | 0.8263 | 0.4442 | 0.8263 | 0.4442 | 0.7354 | 502 |
| | RDPIN | 0.4708 | 0.8346 | 0.4708 | 0.8346 | 0.4708 | 0.7480 | 532 |
| | MR-PIN | 0.5177 | 0.8493 | 0.5177 | 0.8493 | 0.5177 | 0.7703 | 585 |
| PR | SPIN | 0.4274 | 0.8211 | 0.4274 | 0.8211 | 0.4274 | 0.7273 | 483 |
| | DPIN | 0.4363 | 0.8238 | 0.4363 | 0.8238 | 0.4363 | 0.7316 | 493 |
| | RDPIN | 0.4602 | 0.8313 | 0.4602 | 0.8313 | 0.4602 | 0.7429 | 520 |
| | MR-PIN | 0.5611 | 0.8628 | 0.5611 | 0.8628 | 0.5611 | 0.7910 | 634 |
| BC | SPIN | 0.3885 | 0.8089 | 0.3885 | 0.8089 | 0.3885 | 0.7088 | 439 |
| | DPIN | 0.3858 | 0.8081 | 0.3858 | 0.8081 | 0.3858 | 0.7075 | 436 |
| | RDPIN | 0.4522 | 0.8288 | 0.4522 | 0.8288 | 0.4522 | 0.7391 | 511 |
| | MR-PIN | 0.5018 | 0.8443 | 0.5018 | 0.8443 | 0.5018 | 0.7627 | 567 |
| CC | SPIN | 0.3858 | 0.8081 | 0.3858 | 0.8081 | 0.3858 | 0.7075 | 436 |
| | DPIN | 0.3885 | 0.8089 | 0.3885 | 0.8089 | 0.3885 | 0.7088 | 439 |
| | RDPIN | 0.4779 | 0.8368 | 0.4779 | 0.8368 | 0.4779 | 0.7514 | 540 |
| | MR-PIN | 0.5575 | 0.8617 | 0.5575 | 0.8617 | 0.5575 | 0.7893 | 630 |
| TP | SPIN | 0.4159 | 0.8175 | 0.4159 | 0.8175 | 0.4159 | 0.7219 | 470 |
| | DPIN | 0.4389 | 0.8247 | 0.4389 | 0.8247 | 0.4389 | 0.7328 | 496 |
| | RDPIN | 0.4920 | 0.8413 | 0.4920 | 0.8413 | 0.4920 | 0.7581 | 556 |
| | MR-PIN | 0.5602 | 0.8626 | 0.5602 | 0.8626 | 0.5602 | 0.7906 | 633 |
| CLC | SPIN | 0.4204 | 0.8189 | 0.4204 | 0.8189 | 0.4204 | 0.7240 | 475 |
| | DPIN | 0.4664 | 0.8332 | 0.4664 | 0.8332 | 0.4664 | 0.7459 | 527 |
| | RDPIN | 0.4814 | 0.8379 | 0.4814 | 0.8379 | 0.4814 | 0.7531 | 544 |
| | MR-PIN | 0.5212 | 0.8504 | 0.5212 | 0.8504 | 0.5212 | 0.7720 | 589 |
| PeC | SPIN | 0.4133 | 0.8166 | 0.4133 | 0.8166 | 0.4133 | 0.7206 | 467 |
| | DPIN | 0.4150 | 0.8172 | 0.4150 | 0.8172 | 0.4150 | 0.7214 | 469 |
| | RDPIN | 0.4451 | 0.8266 | 0.4451 | 0.8266 | 0.4451 | 0.7358 | 503 |
| | MR-PIN | 0.4752 | 0.8360 | 0.4752 | 0.8360 | 0.4752 | 0.7501 | 537 |
| WDC | SPIN | 0.4735 | 0.8355 | 0.4735 | 0.8355 | 0.4735 | 0.7493 | 535 |
| | DPIN | 0.4708 | 0.8346 | 0.4708 | 0.8346 | 0.4708 | 0.7480 | 532 |
| | RDPIN | 0.5009 | 0.8440 | 0.5009 | 0.8440 | 0.5009 | 0.7623 | 566 |
| | MR-PIN | 0.5239 | 0.8512 | 0.5239 | 0.8512 | 0.5239 | 0.7733 | 592 |
| LBCC | SPIN | 0.4885 | 0.8402 | 0.4885 | 0.8402 | 0.4885 | 0.7564 | 552 |
| | DPIN | 0.4814 | 0.8379 | 0.4814 | 0.8379 | 0.4814 | 0.7531 | 544 |
| | RDPIN | 0.4522 | 0.8288 | 0.4522 | 0.8288 | 0.4522 | 0.7391 | 511 |
| | MR-PIN | 0.5088 | 0.8465 | 0.5088 | 0.8465 | 0.5088 | 0.7661 | 575 |
| UC | SPIN | 0.4690 | 0.8341 | 0.4690 | 0.8341 | 0.4690 | 0.7472 | 530 |
| | DPIN | 0.4858 | 0.8393 | 0.4858 | 0.8393 | 0.4858 | 0.7552 | 549 |
| | RDPIN | 0.4761 | 0.8363 | 0.4761 | 0.8363 | 0.4761 | 0.7505 | 538 |
| | MR-PIN | 0.5221 | 0.8507 | 0.5221 | 0.8507 | 0.5221 | 0.7724 | 590 |
| JDC | SPIN | 0.4832 | 0.8385 | 0.4832 | 0.8385 | 0.4832 | 0.7539 | 546 |
| | DPIN | 0.4876 | 0.8399 | 0.4876 | 0.8399 | 0.4876 | 0.7560 | 551 |
| | RDPIN | 0.4832 | 0.8385 | 0.4832 | 0.8385 | 0.4832 | 0.7539 | 546 |
| | MR-PIN | 0.5195 | 0.8498 | 0.5195 | 0.8498 | 0.5195 | 0.7712 | 587 |

### 3.2.6 Selection and analysis of thresholds

First, we listed the thresholds used by 16 node importance ranking methods in the above experiments on the DIP dataset, as shown in Table 4. Under these thresholds, the overall performances of these methods are optimal. Secondly, in order to study the influence of threshold change on the identification accuracy of essential proteins of the node ranking method, Figure 5 was drawn. As shown in Figure 5(a), we selected and fixed a better value for $p$, that is, when $p$=3.0, each node ranking method reached the maximum value of PRAUC at $q \in [20, 80]$. When p>80 and p<20, the identification accuracies of node ranking methods gradually declined. Similarly, as shown in Figure 5(b), we selected and fixed a better value for $q$, that is, when $q$=60, each node ranking method reached the maximum value of PRAUC at $p\in$ [1.5, 3]. In addition, since the thresholds $p$ and $q$ only control the change of the topological structure of the two-layer network on the MR-PIN, some node ranking methods (such as LBCC method) in Figure 5 are not sensitive to the change of the thresholds, which is the reason why the change of their identification accuracies are not obvious. To sum up, we recommend the range of threshold selection as follows: $p\in$[1.5, 3], $q\in$[20, 80].

Table 3. Performance comparison of 10 network-based centrality methods on four once-refinement subnetworks and their combinations. (top600 / top1130 / PRAUC/ROCAUC)

| Model | LAC | NC | DMNC | LID | DC | EC | PR | CC | TP | CLC |
|---|---|---|---|---|---|---|---|---|---|---|
| $G_T$ | 229/320/ 0.556/ 0.565 | 230/321/ 0.555/ 0.565 | 228/318/ 0.552/ 0.565 | 362/481/ 0.527/ 0.644 | 300/380/ 0.475/ 0.598 | 283/375/ 0.487/ 0.592 | 299/380/ 0.452/ 0.596 | 302/380/ 0.477/ 0.599 | 306/380/ 0.478/ 0.599 | 342/481/ 0.461/ 0.638 |
| $G_S$ | 370/523/ 0.518/ 0.667 | 367/522/ 0.517/ 0.667 | 349/521/ 0.456/ 0.664 | 369/522/ 0.517/ 0.667 | 356/536/ 0.477/ 0.692 | 347/524/ 0.484/ 0.689 | 338/534/ 0.460/ 0.689 | 319/516/ 0.438/ 0.683 | 353/540/ 0.471/ 0.690 | 342/522/ 0.461/ 0.661 |
| $G_M$ | 347/505/ 0.497/ 0.658 | 348/505/ 0.495/ 0.657 | 300/507/ 0.457/ 0.652 | 357/506/ 0.502/ 0.659 | 366/557/ 0.511/ 0.703 | 332/531/ 0.485/ 0.665 | 326/536/ 0.471/ 0.695 | 319/558/ 0.469/ 0.696 | 366/549/ 0.513/ 0.704 | 274/508/ 0.433/ 0.645 |
| $G_C$ | 380/507/ 0.555/ 0.662 | 362/545/ 0.524/ 0.675 | 376/527/ 0.526/ 0.667 | 373/522/ 0.541/ 0.669 | 373/572/ 0.524/ 0.696 | 378/567/ 0.525/ 0.670 | 360/582/ 0.510/ 0.695 | 388/523/ 0.534/ 0.674 | 377/562/ 0.522/ 0.696 | 370/528/ 0.488/ 0.663 |
| $G_T$ $G_S$ | 389/539/ 0.541/ 0.678 | 373/537/ 0.540/ 0.677 | 375/544/ 0.526/ 0.676 | 385/569/ 0.541/ 0.693 | 364/577/ 0.510/ 0.714 | 356/535/ 0.511/ 0.707 | 344/562/ 0.489/ 0.711 | 336/549/ 0.496/ 0.708 | 360/567/ 0.509/ 0.713 | 358/572/ 0.496/ 0.689 |
| $G_T$ $G_M$ | 371/562/ 0.539/ 0.686 | 361/567/ 0.532/ 0.685 | 348/564/ 0.509/ 0.682 | 371/575/ 0.533/ 0.691 | 384/599/ 0.539/ 0.739 | 351/566/ 0.531/ 0.701 | 373/557/ 0.509/ 0.731 | 338/570/ 0.519/ 0.731 | 385/606/ 0.542/ 0.739 | 332/573/ 0.462/ 0.682 |
| $G_T$ $G_C$ | 394/510/ 0.566/ 0.670 | 378/546/ 0.543/ 0.681 | 385/533/ 0.552/ 0.675 | 399/564/ 0.551/ 0.691 | 378/589/ 0.544/ 0.719 | 385/582/ 0.551/ 0.716 | 358/588/ 0.525/ 0.716 | 382/585/ 0.554/ 0.703 | 379/587/ 0.544/ 0.719 | 371/570/ 0.515/ 0.686 |
| $G_S$ $G_M$ | 360/569/ 0.519/ 0.698 | 359/572/ 0.520/ 0.697 | 361/570/ 0.489/ 0.694 | 360/580/ 0.521/ 0.698 | 361/597/ 0.528/ 0.742 | 351/528/ 0.513/ 0.711 | 350/581/ 0.506/ 0.737 | 354/567/ 0.503/ 0.732 | 359/597/ 0.529/ 0.742 | 360/571/ 0.465/ 0.691 |
| $G_S$ $G_C$ | 400/588/ 0.565/ 0.706 | 389/574/ 0.545/ 0.707 | 402/582/ 0.549/ 0.705 | 400/581/ 0.554/ 0.705 | 393/600/ 0.561/ 0.763 | 394/589/ 0.557/ 0.751 | 395/603/ 0.555/ 0.765 | 398/583/ 0.553/ 0.751 | 390/601/ 0.557/ 0.761 | 389/582/ 0.524/ 0.703 |
| $G_M$ $G_C$ | 384/572/ 0.542/ 0.699 | 369/566/ 0.526/ 0.701 | 375/574/ 0.520/ 0.698 | 379/578/ 0.538/ 0.703 | 381/610/ 0.564/ 0.762 | 379/581/ 0.546/ 0.737 | 373/585/ 0.542/ 0.757 | 379/589/ 0.561/ 0.756 | 383/615/ 0.567/ 0.763 | 362/572/ 0.492/ 0.695 |
| $G_T$ $G_S$ $G_M$ | 387/577/ 0.543/ 0.704 | 383/579/ 0.540/ 0.703 | 381/575/ 0.523/ 0.702 | 390/576/ 0.541/ 0.705 | 387/604/ 0.549/ 0.757 | 363/543/ 0.532/ 0.722 | 376/604/ 0.528/ 0.753 | 383/600/ 0.537/ 0.746 | 390/602/ 0.552/ 0.757 | 352/570/ 0.495/ 0.700 |
| $G_T$ $G_S$ $G_C$ | 405/587/ 0.570/ 0.707 | 393/576/ 0.554/ 0.709 | 408/583/ 0.561/ 0.706 | 399/574/ **0.559**/ 0.708 | 401/610/ 0.572/ 0.770 | 406/586/ 0.567/ 0.755 | 385/601/ 0.559/ 0.769 | 407/598/ 0.570/ 0.758 | 401/609/ 0.571/ 0.767 | 403/579/ 0.530/ 0.706 |
| $G_T$ $G_M$ $G_C$ | 404/575/ 0.561/ 0.704 | 388/573/ 0.545/ 0.706 | 408/577/ 0.545/ 0.705 | 389/576/ 0.546/ 0.705 | 404/629/ 0.579/ 0.773 | 378/585/ 0.566/ 0.745 | 388/619/ 0.556/ 0.769 | **408**/595/ 0.580/ 0.768 | 406/624/ 0.582/ 0.774 | 372/573/ 0.509/ 0.700 |
| $G_S$ $G_M$ $G_C$ | 398/582/ 0.555/ 0.711 | 390/592/ 0.543/ 0.710 | 389/584/ 0.539/ 0.710 | 392/584/ 0.549/ 0.712 | 402/633/ 0.581/ 0.778 | **408**/582/ 0.565/ 0.753 | 394/628/ 0.567/ 0.777 | 401/621/ 0.580/ 0.773 | 402/625/ 0.582/ 0.779 | 375/582/ 0.520/ 0.708 |
| $G_T$ $G_S$ $G_M$ $G_C$ (MR-PIN) | **412/587/ 0.568/ 0.714** | **411/592/ 0.554/ 0.713** | **409/588/ 0.557/ 0.713** | **410/597/** 0.557/ **0.712** | **418/633/ 0.592/ 0.785** | 404/**585/ 0.574/ 0.757** | **410/634/ 0.575/ 0.783** | 405/**630/ 0.594/ 0.778** | **418/633/ 0.594/ 0.785** | **408/589/ 0.530/ 0.709** |

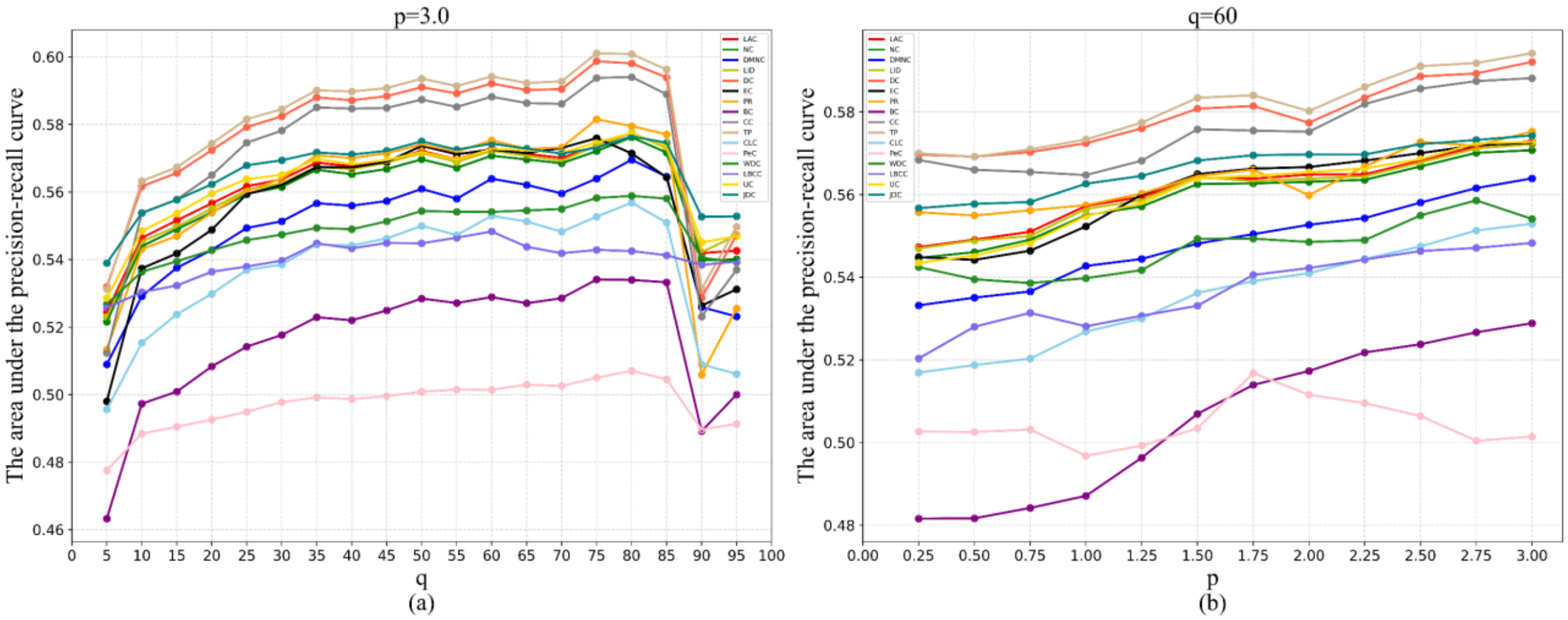

Figure 5. The influence of threshold changes on the performances of 16 node importance ranking methods on DIP dataset.

### 3.2.7 Analysis of reasons for the improvement of identification accuracy of essential proteins

First, we listed the number of interactions, average degree and average clustering coefficient of the four subnetworks in MR-PIN in two datasets, where $p$=1.5, $q$=50. These results showed that the four refined subnetworks obtained by different biological information are existential discrepancy and have different topologies. Secondly, in order to analyze the reason why the proposed method can improve the identification accuracy of essential proteins, we calculated the intersection number of connected essential proteins in each once-refinement subnetwork of the MR-PIN. Although once-refinement network can improve the accuracy of the node importance ranking method to identify essential proteins, some interactions with essential proteins are inevitably filtered during the refining process, resulting in a small number of essential proteins cannot be identified. However, as shown in Figure 6 (p=1.5, q=50), by using the MR-PIN, it is guaranteed that 91.33% of essential proteins can be identified, and the protein importance measurement method can identify different essential proteins in different refinement networks, among which only 29.26% of essential proteins may be jointly identified. This makes it possible for the method in this paper to identify more essential proteins.

## 3.3 Validated by the BioGRID dataset

To verify that the proposed method is applicable to other dataset, we further verified it on a larger protein-protein interaction network (BioGRID dataset). Table 6 compares the performance of 16 node importance ranking methods on the best existing refinement network RDPIN and the MR-PIN obtained by the method in this paper. It can be seen that these methods are better on the MR-PIN, and the improvement effect is very significant. For example, PR method can improve the number of protein identification by 34.85% and 22.36% at top600 and top1190, respectively. The thresholds for each approach at optimum performance also fit within the range we recommended above. Therefore, all these further prove the effectiveness and universality of MR-PIN.

Table 4. Threshold values selected for 16 node importance ranking methods on DIP dataset.

| Methods | The threshold used in this paper | |
|---|---|---|
| | *p* | *q* |
| LAC | 3.0 | 40 |
| NC | 3.0 | 20 |
| DMNC | 3.0 | 35 |
| LID | 1.5 | 30 |
| DC | 3.0 | 60 |
| EC | 3.0 | 50 |
| PR | 3.0 | 60 |
| BC | 3.0 | 75 |
| CC | 3.0 | 75 |
| TP | 3.0 | 60 |
| CLC | 1.5 | 35 |
| PeC | 1.75 | 55 |
| WDC | 2.75 | 35 |
| LBCC | 3.0 | 35 |
| UC | 3.0 | 35 |
| JDC | 1.75 | 80 |

Table 5. Topological characteristics of SPIN and four constructed subnetworks of the MR-PIN on YDIP and BioGRID datasets.

| Datasets | Networks | Interactions | Average degree | Average clustering coefficient |
|---|---|---|---|---|
| | SPIN | 15,166 | 6.3911 | 0.0923 |
| | GT | 5,167 | 2.1774 | 0.0562 |
| YDIP | GS | 5,827 | 2.4555 | 0.0645 |
| | GM | 3,531 | 1.4880 | 0.1268 |
| | GC | 4,360 | 1.8373 | 0.0466 |
| | SPIN | 52,833 | 18.8152 | 0.2250 |
| | GT | 19,591 | 6.9769 | 0.1147 |
| BioGRID | GS | 18,667 | 6.6478 | 0.1011 |
| | GM | 11,105 | 3.9547 | 0.1672 |
| | GC | 16,162 | 5.7557 | 0.0761 |

## 4 Conclusion

In this paper, a multi-layer refined netwok model for essential proteins based on multiple once-refinement networks was proposed. That is, firstly, four once-refinement networks were obtained by using four biological information (gene expression profile, subcellular localization information, protein complex and orthologous information) to form a four-layer heterogeneous network. Then, the node importance ranking method was used to calculate the importance of proteins in each layer of a refinement network. Finally, the geometric mean method was used to comprehensively evaluate the importance of proteins in this heterogeneous network. We used 16 node importance ranking methods (LAC, NC, DMNC, LID, DC, EC, PR, BC, CC, TP, CLC, PeC, WDC, UC, JDC, LBCC) and two protein-protein interaction networks (DIP dataset and BioGRID dataset), respectively verified the identification number of essential proteins at top100-top600, the jackknifing method, accuracy and other six evaluation indicators, ROCAUC as well as PRAUC. The experimental results showed that compared with the existing once-refinement network DPIN and twice-refinement network RDPIN, the MR-PIN is a better model for essential protein identification, which can effectively improve the identification accuracy of essential proteins by each node importance ranking method.

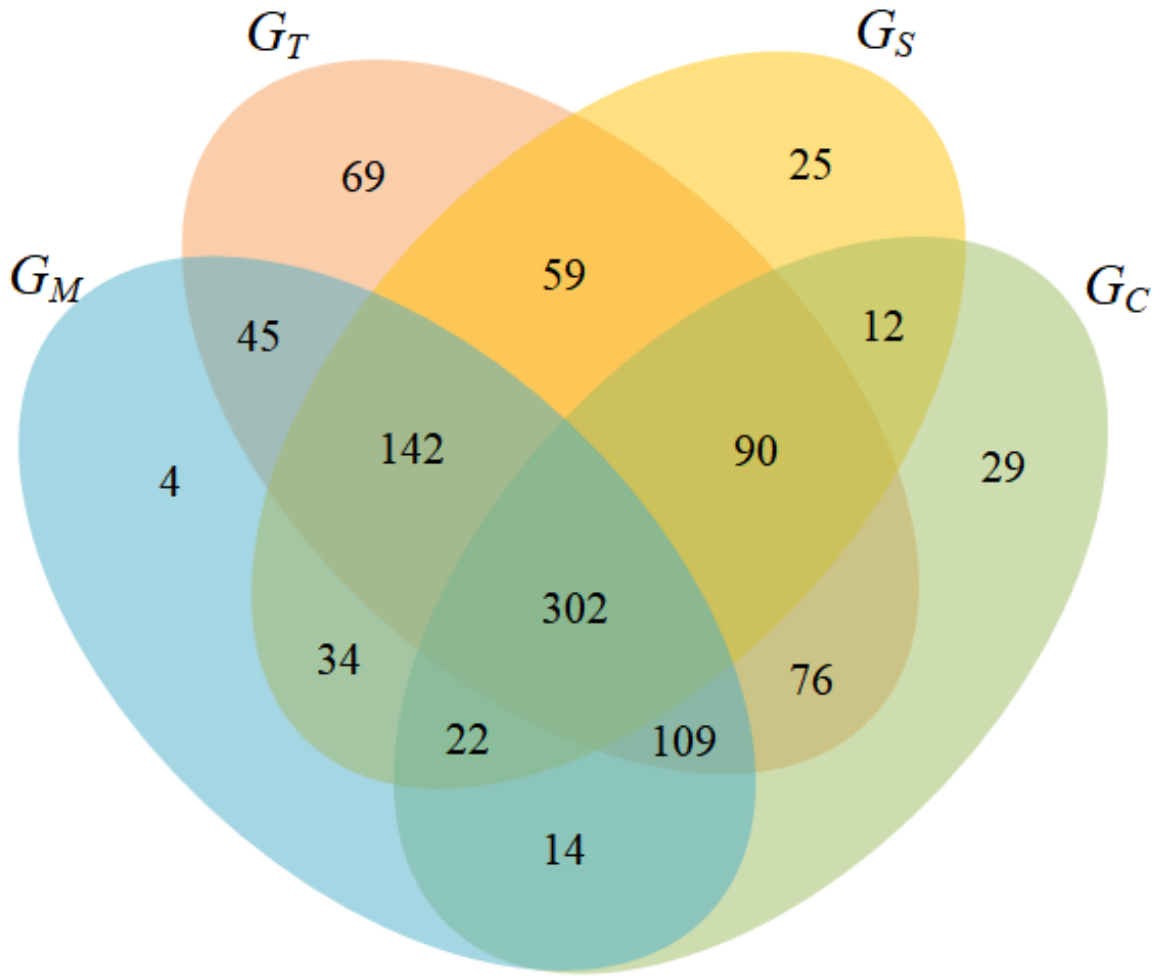

Figure 6. The intersection number of connected essential proteins in four once refinement networks on DIP dataset.

Table 6. Comparison of the performance of 16 node importance ranking methods on RDPIN and MR-PIN and their thresholds on BioGRID dataset.

| Methods | RDPIN | | | | MR-PIN | | | | Thresholds | |
|---|---|---|---|---|---|---|---|---|---|---|
| | top100/top600 /top1130 | ACC | ROC AUC | PR AUC | top100/top600 /top1199 | ACC | ROC AUC | PR AUC | $p$ | $q$ |
| LAC | 57/339/588 | 0.7824 | 0.736 | 0.448 | 88/401/664 | 0.8095 | 0.795 | 0.572 | 3.0 | 35 |
| NC | 57/347/588 | 0.7824 | 0.737 | 0.461 | 80/403/661 | 0.8084 | 0.793 | 0.571 | 3.0 | 45 |
| DMNC | 28/307/540 | 0.7653 | 0.722 | 0.404 | 88/417/651 | 0.8048 | 0.788 | 0.572 | 3.0 | 60 |
| LID | 57/340/600 | 0.7867 | 0.738 | 0.451 | 88/405/668 | 0.8109 | 0.789 | 0.583 | 3.0 | 75 |
| DC | 61/341/596 | 0.7853 | 0.758 | 0.474 | 84/405/675 | 0.8134 | 0.810 | 0.585 | 3.0 | 75 |
| EC | 51/351/559 | 0.7721 | 0.750 | 0.447 | 73/371/621 | 0.7942 | 0.776 | 0.506 | 2.75 | 75 |
| PR | 61/307/550 | 0.7689 | 0.744 | 0.442 | 86/414/673 | 0.8127 | 0.809 | 0.576 | 3.0 | 45 |
| BC | 41/260/490 | 0.7475 | 0.715 | 0.379 | 87/364/617 | 0.7927 | 0.778 | 0.537 | 3.0 | 75 |
| CC | 50/273/492 | 0.7482 | 0.729 | 0.398 | 88/403/622 | 0.7945 | 0.796 | 0.560 | 3.0 | 75 |
| TP | 70/339/537 | 0.7642 | 0.748 | 0.463 | 87/406/664 | 0.8095 | 0.809 | 0.574 | 3.0 | 60 |
| CLC | 22/226/480 | 0.7439 | 0.705 | 0.361 | 87/411/651 | 0.8048 | 0.784 | 0.565 | 3.0 | 75 |
| PeC | 63/327/548 | 0.7682 | 0.719 | 0.430 | 68/338/551 | 0.7692 | 0.710 | 0.446 | 1.5 | 75 |
| WDC | 60/347/588 | 0.7824 | 0.748 | 0.465 | 79/379/636 | 0.7995 | 0.784 | 0.545 | 3.0 | 55 |
| UC | 31/355/603 | 0.7877 | 0.741 | 0.456 | 81/407/650 | 0.8045 | 0.794 | 0.572 | 3.0 | 50 |
| JDC | 72/342/594 | 0.7845 | 0.737 | 0.484 | 87/408/659 | 0.8077 | 0.788 | 0.575 | 3.0 | 45 |
| LBCC | 17/314/588 | 0.7824 | 0.693 | 0.409 | 54/363/618 | 0.7931 | 0.719 | 0.474 | 2.75 | 35 |

## Acknowledgment

This work was supported in part by the Scientific Research Fund of Hunan Provincial Education Department of China under Grant 19B231, in part by the Hunan Provincial Postgraduate Research and Innovation Foundation Project of China under Grant CX20211183. Haoyue Wang and Li Pan contributed equally to this work.

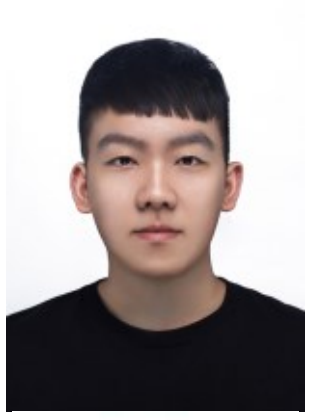

**HAOYUE WANG** is currently pursuing the Graduate degree with the Hunan Institute of Science and Technology. His research interests include bioinformatics and complex networks.

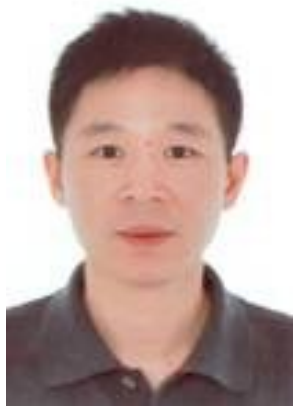

**LI PAN** received the Ph.D. degree in computer science from Tongji University, China, in 2009. He is currently a Professor with the School of Information Science and Engineering, Hunan Institute of Science and Technology. His research interests include bioinformatics and Petri nets.

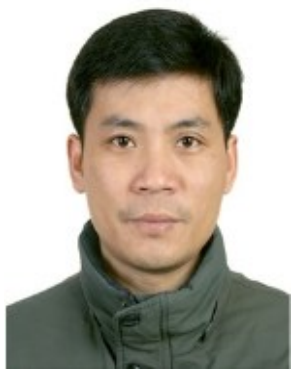

**BO YANG** received the B.Sc. degree in mechanical engineering from Zhengzhou University, China, in 1996, the M.Sc. degree in computer application technology from Xiangtan University, China, in 2004, and the Ph.D. degree in mechanical and electrical engineering from Central South University, China, in 2010. Since 2012, he has been an Associate Professor with the College of Information Science and Technology, Hunan Institute of Science and Technology. His research interests include MR brain image analysis, statistical pattern recognition, and machine learning

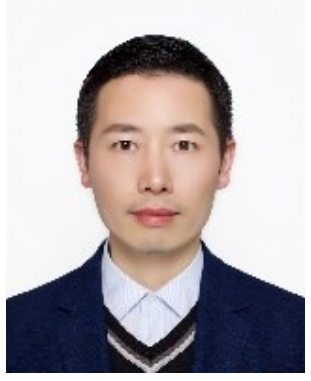

**JUNQIANG JIANG** received the Ph.D. degree in software engineering from Hunan University, Changsha, China, in 2017. He is currently an Associate Professor in the School of Information Science and Engineering, Hunan Institute of Science and Technology, China. His main research interests include cloud computing, parallel computing and workflow scheduling. He is a member of IEEE. He has been awarded the young elite teacher of Hunan province general universities. His research has been supported by the natural science foundation of Hunan province, and the scientific research fund of Hunan provincial education department.

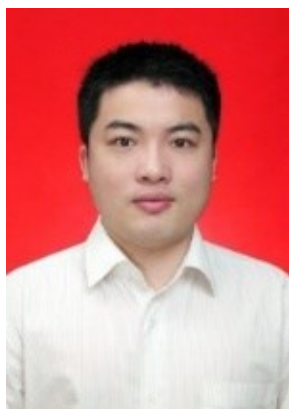

**WENBIN LI** received the B.S. degree in computer science and technology from Hunan Normal University, Changsha, China, in 2003, the M.S. degree in computer applications technology from the Changsha University of Science and Technology, Changsha, in 2006, and the Ph.D. degree in control engineering from Central South University, Changsha, in 2020. His research interests include industrial process control, evolutionary computation, and multi-objective optimization.